\documentclass[review,12pt]{elsarticle}
    \usepackage{amsmath}
    \usepackage{dsfont}
    \usepackage{graphicx}
    \usepackage{amsfonts}
    \usepackage{graphicx}
    \usepackage{amssymb}
    \usepackage{color}
    \usepackage{amsthm,bm}
    \usepackage{pgf,pgfarrows}
    \usepackage[latin1]{inputenc}
    \usepackage{amssymb}

\def\bfA{{\mathbf{A}}}
\def\bfY{{\mathbf{Y}}}
\def\bfS{{\mathbf{S}}}

\newcommand{\figwidth}{0.8\columnwidth}
\newcommand{\figext}{eps}

\journal{Ultramicroscopy}

\begin{document}

\begin{frontmatter}

\title{Spectral mixture analysis of EELS spectrum-images}

\author[label1]{Nicolas Dobigeon}
\ead{Nicolas.Dobigeon@enseeiht.fr}
\author[label2]{Nathalie Brun\corref{cor1}}
\ead{nathalie.brun@u-psud.fr}

\address[label1]{University of Toulouse, IRIT/INP-ENSEEIHT, 2 rue Camichel, 31071 Toulouse Cedex 7, France}
\address[label2]{University of Paris Sud, Laboratoire de Physique des Solides, CNRS, UMR 8502, 91405 Orsay Cedex, France}
\cortext[cor1]{Corresponding author.}

\begin{abstract}
Recent advances in detectors and computer science have enabled the
acquisition and the processing of multidimensional datasets, in
particular in the field of spectral imaging. Benefiting from these
new developments, earth scientists try to recover the reflectance
spectra of macroscopic materials (e.g., water, grass, mineral
types...) present in an observed scene and to estimate their
respective proportions in each mixed pixel of the acquired image.
This task is usually referred to as \emph{spectral mixture analysis}
or \emph{spectral unmixing} (SU). SU aims at decomposing the
measured pixel spectrum into a collection of constituent spectra,
called \emph{endmembers}, and a set of corresponding fractions
(\emph{abundances}) that indicate the proportion of each endmember
present in the pixel. Similarly, when processing spectrum-images,
microscopists usually try to map elemental, physical and chemical
state information of a given material. This paper reports how a SU
algorithm dedicated to remote sensing hyperspectral images can be
successfully applied to analyze spectrum-image resulting from
electron energy-loss spectroscopy (EELS). SU generally overcomes
standard limitations inherent to other multivariate statistical
analysis methods, such as \emph{principal component analysis} (PCA)
or \emph{independent component analysis} (ICA), that have been
previously used to analyze EELS maps. Indeed, ICA and PCA may
perform poorly for linear spectral mixture analysis due to the
strong dependence between the abundances of the different materials.
One example is presented here to demonstrate the potential of this
technique for EELS analysis.
\end{abstract}


\end{frontmatter}

% \linenumbers

\section{Introduction}

Over the two last decades, scanning transmission electron microscopy
(STEM) have benefit from important advances in electron-based
instrumentation and technology. These recent advances have enabled
the development of electron energy-loss spectroscopy (EELS). EELS
provide spectrum-images, that have been widely used in various
applications, including material science and chemical analysis
\cite{Colliex1994} \cite{Pennycook2011}. The multidimensional data
coming from EELS analysis exploit inherent spatial information to
build elemental maps. An elemental map is useful per se, however it
does not exploit additional crucial information present in the
acquired spectrum image. As EELS signal is sensitive to chemical
changes and atom environment, building a map of the different
materials would be more much more relevant. Therefore, there is a
real need for efficient techniques to process EELS spectrum-images,
able to identify and quantify the spectral components that represent
the different compounds present in the imaged sample.

Attempts to extract information from EELS spectra were conducted in
1999 mainly based on multivariate data analysis techniques,
specifically principal component analysis (PCA) \cite{Bonnet1999}. A
PCA-based method was written for DigitalMicrograph and
commercialized by Ishizuka in 2001 \cite{HREM} and is now rather
widely used for data filtering and dimensional reduction
\cite{Bosman2006}. However, such analysis faces the difficulty of
extracting physically meaningful spectra from the computed
eigenvalues.

Conversely, independent component analysis (ICA) aims at identifying
statistically independent components from multivariate data. In
2005, Bonnet and Nuzillard \cite{Bonnet2005} applied the ICA-based
SOBI algorithm to process spectrum image data set. The authors
noticed that, since EELS spectra are not composed of separated
peaks, the independence hypothesis is not fulfilled. To overcome
this issue, successive derivatives of EELS-spectra are analyzed.
From this analysis, it seems that first derivatives produce more
interpretable results than second derivatives. Unfortunately, this
finding was empirical and no theoretical argument was found to
justify this point. De La Peña proposed in \cite{delapena2011} to
use a kernelized version of ICA. This approach allows C, SnO2 and
TiO2 signals to be successfully separated while analyzing a
spinodally decomposed solid solution. Satisfactory quantitative
analysis was obtained but no fine structure analysis was performed.
The authors noticed that difficulties could be encountered because
of multiple scattering and energy instabilities introducing non
linearity.

Recently a matrix factorization technique has been proposed to map
plasmon modes on silver nanorods \cite{Guiton2011}. The analysis,
relying on the software AXSIA developed by Keenan and co-workers
\cite{Keenan2009} consists in looking for a rotation matrix to be
applied on orthogonal factors to maximize the intrinsic
``simplicity'' of the decomposition. Specifically, the optimal
solution is defined by the sparsity of the spatial distribution of
each individual material.

In a significantly different area -- namely remote sensing and
geoscience -- reflectance spectroscopy is widely used to
characterize and discriminate materials on the Earth surface for
various applications \cite{Keshava2002}. Usually mounted on
aircrafts, balloons or satellites, spectral sensors collect
electromagnetic radiations from the Earth surface. Most of the
recorded signals are reflectance spectroscopic signals measured in
the infra-red/visible range. The collection of these signals over an
observed scene provides a multi-band image formed as a 3-dimensional
data cube. Each pixel of the atmospheric-corrected image is
characterized by a vector of reflectance measurements. Specifically,
hyperspectral images are composed of pixels with several hundreds of
narrow and contiguous spectral bands.

Faced with this amount of data, the geophysicist community has
developed analysis methods to extract physical information from
these images. One of the main objectives of these methods is to
identify spectral properties corresponding to distinct materials in
a given scene and thus to get classification maps of the image
pixels. However, because of the intrinsically limited spatial
resolution of the hyperspectral sensors, several materials (e.g.,
water, grass, mineral types...) usually contribute to the spectrum
measured at a given single pixel. The resulting spectral measurement
is a combination of the individual spectra that are characteristic
of the macroscopic materials. Consequently, techniques to estimate
the constituent substance spectra and their respective proportions
from mixed pixels are needed. Spectral unmixing is the procedure
that aims at i) decomposing the measured pixel spectrum into a
collection of constituent spectra, or endmembers, and ii) estimating
the corresponding fractions, or abundances, that indicate the
proportion of each endmember present in the pixel
\cite{Bioucas2012jstars}.

What is usually known as ``spectrum image'' in the microscopist
community corresponds very precisely to a ``hyperspectral image''
for the geoscience-related applications. The analogy between these
two fields of research is undeniable. However, at the present time
microscopists are less advanced in their ability of conducting
efficient multivariate analysis of their data. In this work we
describe how a recent spectral unmixing algorithm developed by
Dobigeon et al. \cite{Dobigeon2009} for analyzing hyperspectral
images can be successfully applied to spectrum images resulting from
EELS maps.

\section{Methods and experimental setup}
\subsection{Spectral mixture analysis}
\label{subsec:BLU} 
\begin{figure}[h!]
\centering
\includegraphics[width=\figwidth]{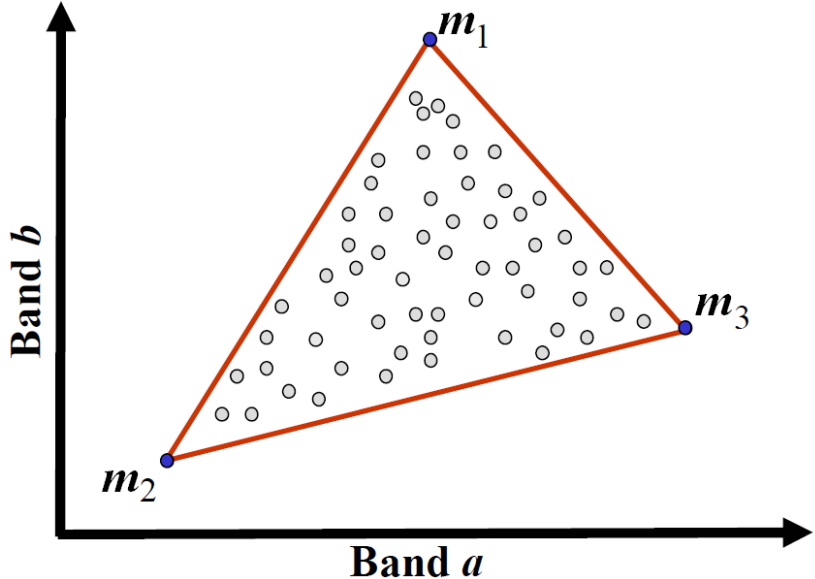}
\caption{Geometrical formulation of spectral mixture analysis (SMA).
The scatterplot represents the data observed in a 2-D space. The
mixed pixels (grey circles) belong to the simplex (simplest
geometric figure that is not degenerate in n-dimensions), whose
vertices are the 3 endmembers. SMA algorithms exploit different
properties of the simplex. \label{fig:fig_1}}
\end{figure}

This paragraph formulates the so-called spectral
unmixing or spectral mixture analysis. Let $\bfY$ denote the $L$ by
$N$ observed data matrix that gathers the whole set of $N$ measured
pixel spectra. Each column of the $\bfY$ is a vector of size $L$
which corresponds to the reflectances measured in the $L$ spectral
bands. The spectral mixture analysis (SMA) conducted on the spectrum
image consists of decomposing this matrix $\bfY$ into a product
matrix $\bfS\bfA$. In this decomposition scheme, each column of the
$L$ by $R$ matrix $\bfS$ is the spectral signature of a constituent
(endmember). Conversely, each column of $\bfA$ is a set of $R$
coefficients corresponding to the relative proportions of the
signatures in the pixels. Thus, like any factorization matrix
method, SMA estimates the two latent variables $\bfS$ and $\bfA$
leading to the product $\bfS\bfA$ that best approximates the
observed matrix $\bfY$. Since this decomposition is non-unique, the
problem of estimating $\bfS$ and $\bfA$ from $\bfY$ is
ill-conditioned. To reduce the set of admissible solutions,
additional constraints on $\bfS$ and $\bfA$ are considered. First,
as any non-negative matrix factorization (NMF) approach, the
elements of $\bfS$ and $\bfA$ are assumed to be positive. Moreover,
since the coefficients in each column of $\bfA$ represent
proportions, it is natural to consider an additional sum-to-one
constraint on these columns. This constrained matrix factorization
problem has been widely addressed in the geoscience and remote
sensing literature since SMA is a crucial step in analyzing
multi-band images, e.g., hyperspectral data. Note that, from a
geometrical point of view, SMA consists of identifying the vertices
of a lower dimensional simplex formed by the observed data (Fig.
\ref{fig:fig_1}). Indeed, under the positivity and additivity
constraints introduced previously, the observed spectral vectors
form a simplex whose vertices correspond to the endmembers to be
identified. $R+1$ pure endmembers spectra form the vertices of an
$R$-simplex,. Thus, as examples, a 2-simplex is a triangle (Fig.
\ref{fig:fig_1}), a 3-simplex a tetrahedron... Several algorithms of
the geoscience and remote sensing literature have proposed to
exploit this geometrical formulation to solve the spectral unmixing
problem. Vertex Component Analysis (VCA) is one of the most popular
geometric algorithm \cite{Nascimento2005}. It consists of
iteratively i) projecting the data onto the direction orthogonal to
the subspace spanned by the endmembers previously identified ii)
assigning the extreme projection as a new endmember.

Geometrical algorithms have the great advantage of being
computationally efficient. However, most of them, such as VCA, rely
on the hard hypothesis of ``pure pixels'', i.e., they assume that
the endmembers are present among the observed pixels. Unfortunately,
this assumption can be rarely ensured and alternative strategies
must be considered. In this work, SMA is conducted with the Bayesian
Linear Unmixing (BLU) method proposed by Dobigeon et al.
\cite{Dobigeon2009}. Originally developed to address the
hyperspectral unmixing of remote sensing images, BLU relies on a
Bayesian formulation of the estimation problem. This Bayesian
framework allows the positivity and sum-to-one constraints
introduced above to be conveniently included into the observation
model. The proposed BLU method has the great advantage of recovering
the endmember signatures S and their respective proportions A
jointly in a single step. Naturally, this strategy casts SMA as a
standard blind source separation (BSS) problem. Moreover, contrary
to geometrical based algorithm like VCA, it does not require the
assumption of having pure pixels among the data. Moreover, note that
BLU solves the endmember estimation problem directly on a lower
dimensional space, exploiting the intrinsic geometrical
interpretation of SMA noticed above. By conducting SMA in the
subspace spanned by the identified simplex, the number of freedom
associated with the parameters to be estimated is significantly
reduced when compared to other algorithms dedicated to SMA. The
methodology of BLU can be summarized as follows. First, appropriate
prior distributions $p(\bfS)$ and $p(\bfA)$ are assigned to the
unknown parameters $\bfS$ and $\bfA$, respectively. These
distributions are chosen to ensure the positivity and sum-to-one
constraints imposed on the unknown matrices $\bfS$ and $\bfA$. Then,
based on this prior modeling and the well-admitted assumption of a
Gaussian likelihood $p(\bfY|\bfS, \bfA)$, the joint posterior
distribution $p(\bfS, \bfA|\bfY)$ is computed using the Bayes
paradigm. Unfortunately, this posterior is too complex to easily
derive the closed-form expressions of the standard Bayesian
estimators, such as the maximum a posteriori or posterior mean.
Consequently, a Markov chain Monte Carlo (MCMC) algorithm is
designed to generate samples $\bfS^{(t)}$ and $\bfA^{(t)}$
($t=1,...N_{MC}$) asymptotically distributed according to the
posterior of interest. Finally, the Bayesian estimators of the
endmember matrix $\bfS$ and the proportion matrix $\bfA$ are then
approximated using these $N_{MC}$ generated samples. Note that a
Matlab$^{\copyright}$ code of the BLU algorithm is freely available
online \cite{Dobigeon2009}.

\subsection{Experimental data}

In the following sections, SMA of a spectrum-image of nanoparticles
is conducted. More precisely, the ability of BLU to provide
interpretable spectral signatures is demonstrated, thus overcoming
the standard limitations inherent to other multivariate analysis
techniques, such as PCA and ICA.

\begin{figure}[h!] \centering
\includegraphics[width=\figwidth]{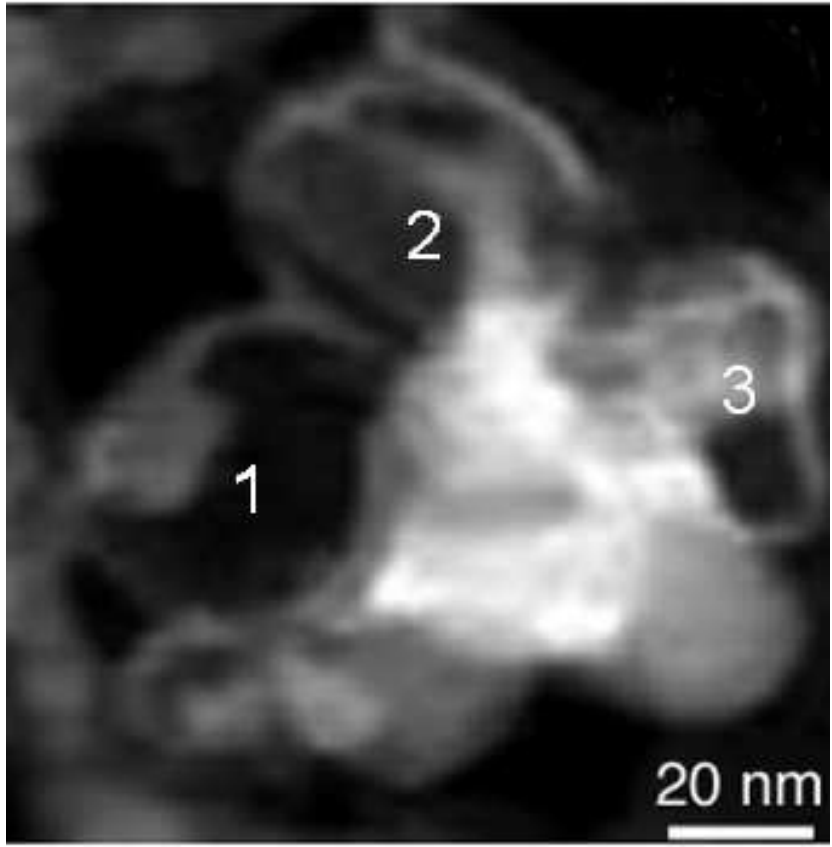}
\caption{HADF image corresponding to a $64\times64$ spectrum-image
recorded in an area rich in nanoparticles containing boron (pure
boron, boron oxide or h-BN).\label{fig:fig_2}}
\end{figure}

\begin{figure}[h!]
\centering
\includegraphics[width=\figwidth]{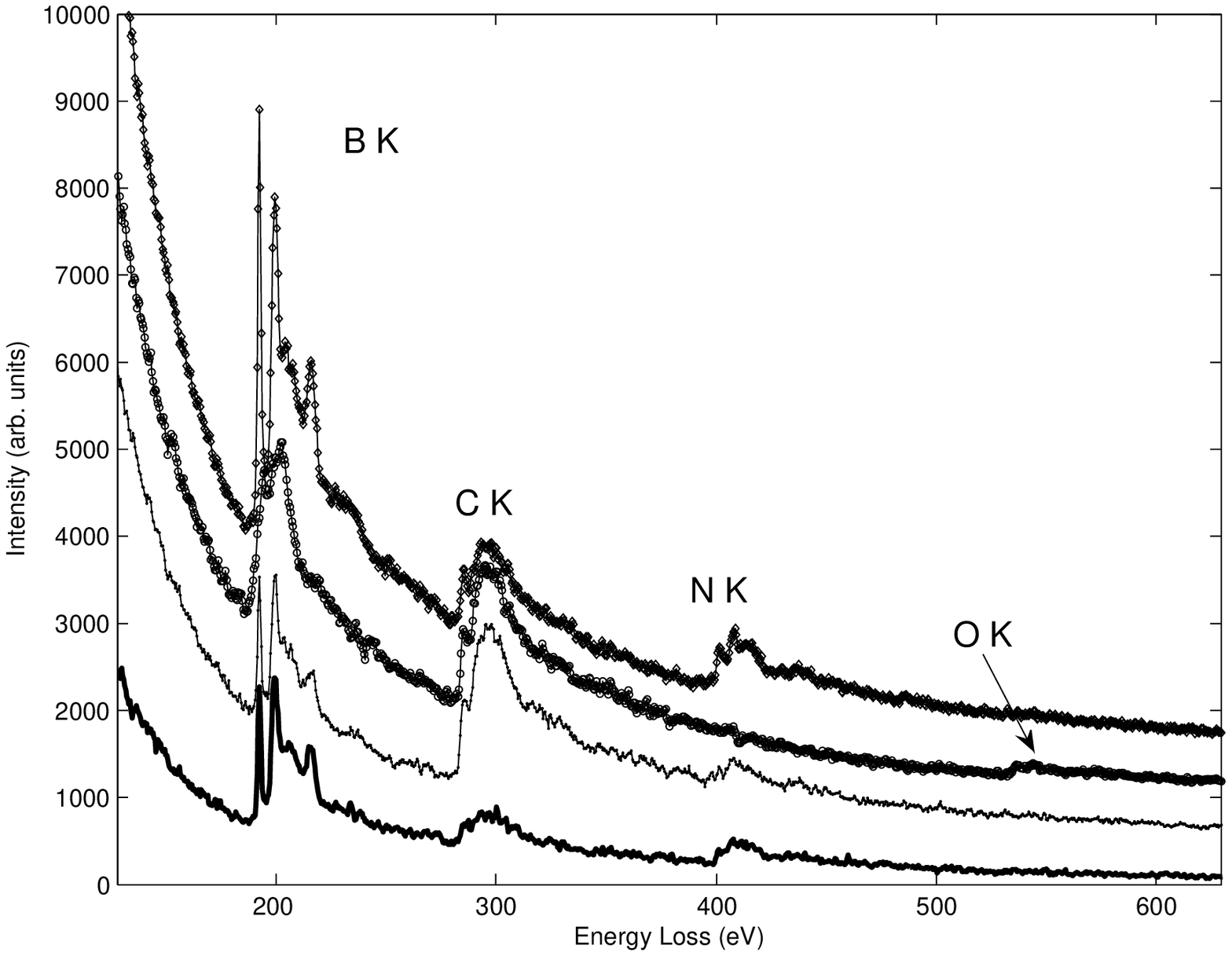}
\caption{Typical EELS spectra extracted from the $64\times64$
spectrum image of Fig. \ref{fig:fig_2}. Boron, carbon, nitrogen and
oxygen K edges are represented.\label{fig:fig_3} }
\end{figure}

The analyzed dataset consists of a $64\times64$ pixel spectrum-image
acquired in 1340 energy channels over a region composed of several
nanocages in a boron-nitride nanotubes (BNNT) sample. Note that
nanocages are supported by a holey carbon film for TEM analysis.
These data have been extensively described and analyzed in
\cite{Arenal2008} and a high angle dark field image of the region of
interest is depicted in Fig. \ref{fig:fig_2}. In this study, ELNES
``fingerprints'' for different bonding configurations of boron (B-B,
B-O, B-N$^{\pi*}$, B-N$^{\sigma*}$) have been extracted from
selected area of the sample. Then reconstructed spectra are computed
according to a linear combination of a power law and four
fingerprints thanks to a multiple least squares fitting procedure.
Fig. \ref{fig:fig_3} displays characteristic spectra with the
involved edges (B-K, C-K, N-K and O-K).

\section{Results}

\subsection{Principal component analysis}

PCA has demonstrated its ability to extract relevant information
from multidimensional data. For instance, this method and its
application to EELS data have been described in \cite{Bonnet1999}.
Moreover, this powerful multivariate analysis technique is also able
to provide a minimal representation of the signal of interest,
performing an explicit dimensionality reduction. In particular, in
the specific context of SMA and according to the geometrical
interpretation of spectral unmixing given in the previous section,
the intrinsic dimension of the data is straight related with the
number of endmembers to be recovered. When the mixed pixels are
assumed to be obtained from the constrained linear combination of
$R$ spectral components, only $R-1$ dimensions are required to
represent the data without loss of any information.

The method commonly advocated to determine the intrinsic dimensional
of the data is to monitor the eigenvalues obtained by PCA. Only
eigenvectors associated with eigenvalues of highest magnitudes are
retained as significant contributions. Several criteria have been
proposed to decide on the number of relevant eigenvalues. One
solution consists in plotting the logarithm of these eigenvalues
previously arranged in decreasing order. Ideally, smallest values
related to noise correspond to the final linear part of the plot.
However, the actual dimensionality of the data is generally
difficult to assessed in practice, since changes between two
adjacent eigenvalues may not be significant. This is typically the
case for real data encountered in hyperspectral imagery, such as the
HYDICE image scene. In \cite{Chang2004tgrs}, the authors conclude
that only a crude estimate of the number of signal sources can be
provided. Indeed the signature of an unique target may vary
significantly from one area to another. Moreover, signal of weak
amplitude might be difficult to separate from noise.

\begin{figure}[h!]
\centering
\includegraphics[width=\figwidth]{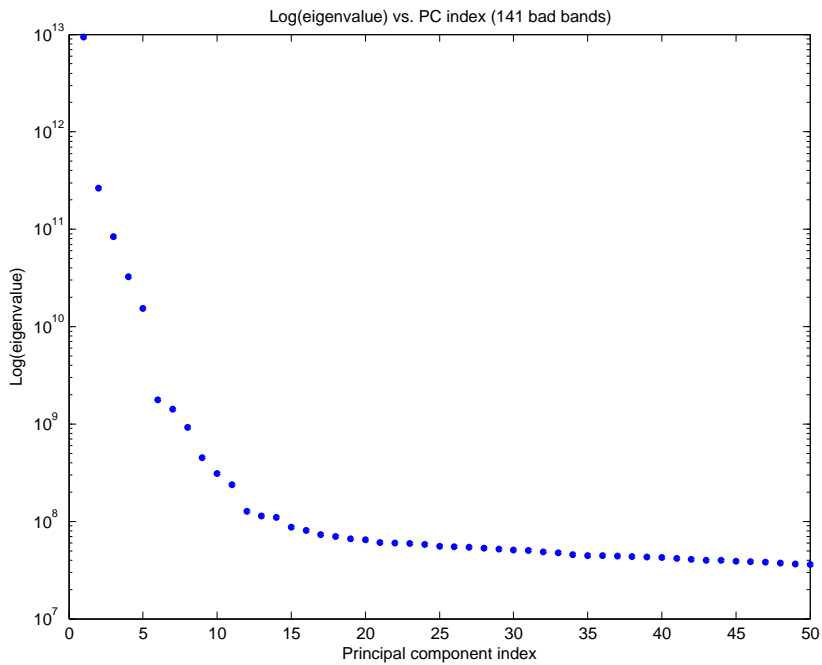}
\caption{Principal component eigenvalues for the analyzed
spectrum-image arranged in a decreasing order (depicted in
logarithmic scale). The threshold between eigenvalues associated
signal and those associated with noise is not easy to determine.
Indeed, there is no drastic drop in the eigenvalue distribution. In
this work, the intrinsic dimensionality of the data is estimated
around 7.\label{fig:fig_4} }
\end{figure}

The eigenvalues for the analyzed spectrum-image are plotted in Fig.
\ref{fig:fig_4}. As expected, the threshold can not be clearly
defined since there is no drastic drop in the eigenvalues
distribution. The main objective of the study is to separate
B-N$^{\pi*}$ from B-N$^{\sigma*}$ while keeping a minimum number of
components for the other signatures. In practice, the analysis of
the considered EELS dataset has been conducted with a number of
spectral signatures $R$ ranging from $6$ to $8$ for each evaluated
analysis method (PCA, ICA and SMA).

\begin{figure}[h!]
\centering
\includegraphics[width=\figwidth]{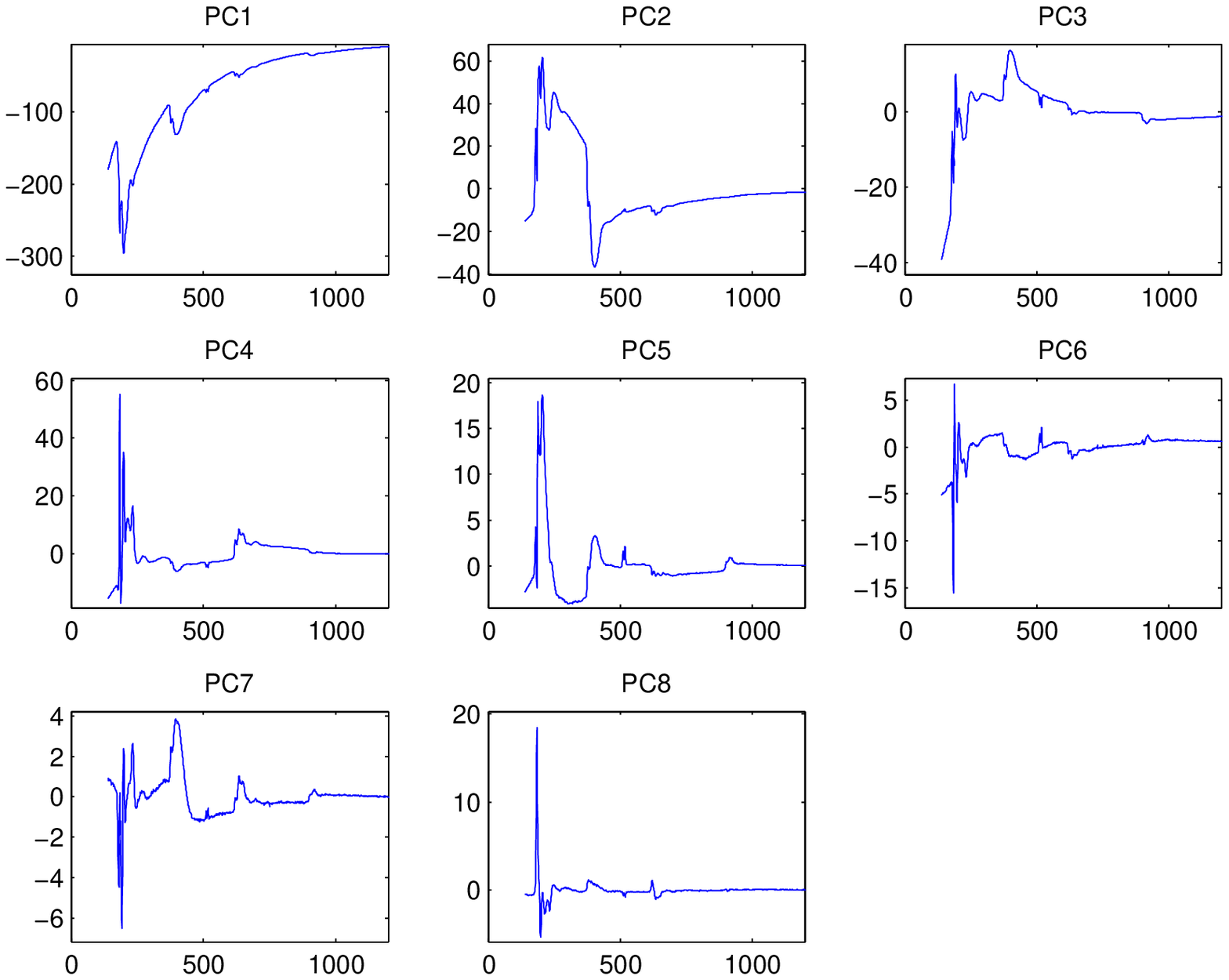}
\caption{First $8$ most relevant components determined by PCA. These
components are orthogonal and thus do not correspond to any
``physically'' significant spectral signature.\label{fig:fig_5} }
\end{figure}

PCA has been performed with $R=8$ using the open source Hyperspy
toolbox \cite{delapena2011}, with a weighted version of PCA. The
first eight spectra corresponding to PCA eigenvectors of highest
relevance are displayed in Fig. \ref{fig:fig_5}. It clearly appears
that these components do not correspond to any meaningful physical
spectra. Consequently, they do not allow any interpretation,
quantification or comparison with reference spectra. This can be
explained by the fact that PCA searches for orthogonal components,
which is not a realistic assumption for EELS application.

\subsection{Independent component analysis}

\begin{figure}[h!]
\centering
\includegraphics[width=\figwidth]{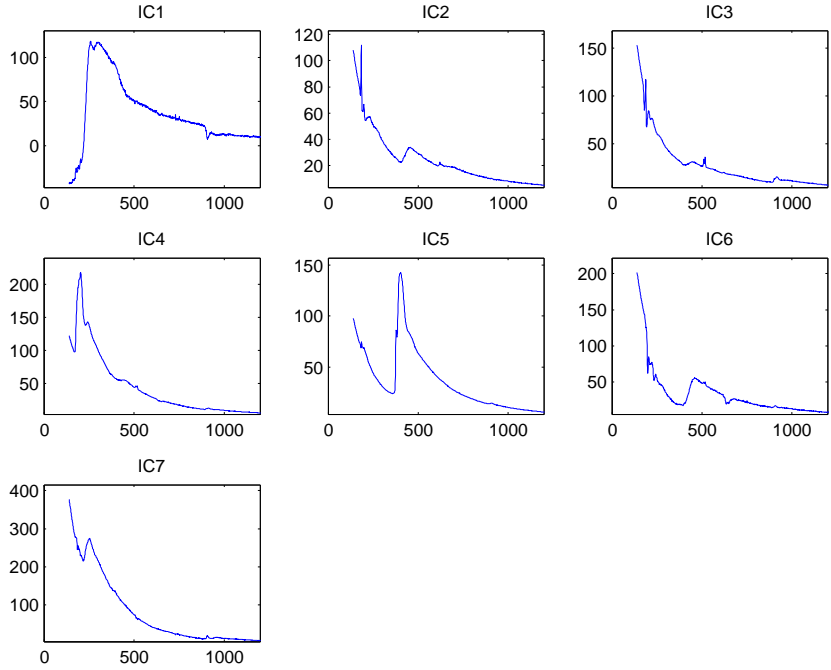}
\caption{Spectral components extracted by CubICA. Some components
are completely different from the reference spectra and have no
physical meaning. The corresponding maps are depicted in Fig.
\ref{fig:fig_7}.\label{fig:fig_6} }
\end{figure}

\begin{figure}[h!]
\centering
\includegraphics[width=\figwidth]{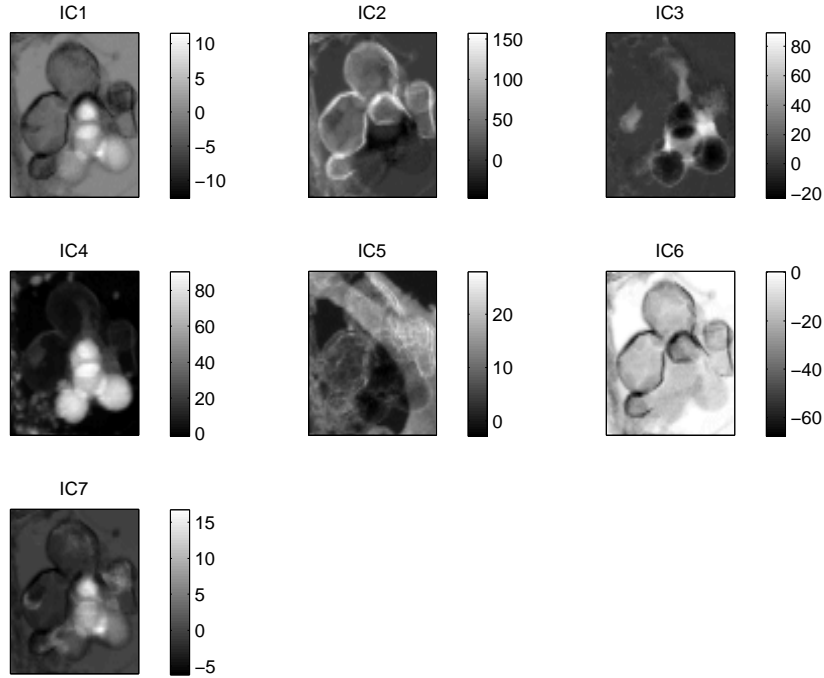}
\caption{Maps of the spectral components extracted by CubICA (the
spectral signatures are depicted in Fig.
\ref{fig:fig_6}).\label{fig:fig_7}}
\end{figure}

Whereas PCA searches for orthogonal components, ICA aims at
identifying statistically independent components. Different measures
of independence have been exploited in the literature, corresponding
to different algorithms. In this work ICA has been performed using
the open source Hyperspy toolbox \cite{delapena2011}, choosing the
CubICA algorithm. In contrast to other ICA methods, CubICA can be
used without any parameter adjustments. It is thus easy to use and
has been already applied for EELS spectrum-imaging data analysis.
After visual expertise of the results obtained for $R=6$, $R=7$ and
$R=8$ components, we considered that $R=7$ provides the most
physically interpretable results. The identified components and
their respective abundance maps are depicted in Fig. \ref{fig:fig_6}
and Fig. \ref{fig:fig_7}, respectively. The component IC5 is clearly
identified as amorphous carbon and the map corresponds to the carbon
supporting film. The 3 components IC1, IC4 and IC7 are associated
with pure Boron and the corresponding abundance maps match those
obtained in \cite{Arenal2008} for this specific compound. The
separation of the signature into $3$ different components may be
explained by thickness effects. By analyzing the abundance map
associated with IC2, this component can be identified as
B-N$^{\pi*}$, but its features are significantly different from
those of the reference spectrum in \cite{Arenal2008} and do not
correspond to any proper EELS edge. Similarly IC3 should correspond
to B2O3. However, whereas O-K edge appears properly, no real
physical edge for the B K is obtained. Finally, unfortunately,
component IC6 does not correspond to physically acceptable spectra
and its abundance map is not interpretable. As a consequence, we
have to conclude that ICA has failed to completely unmix the signal
sources. In particular, we do not obtain the signature for
B-N$^{\pi*}$, B-N$^{\sigma*}$. This limitation of ICA has already
been noticed in \cite{Nascimento2005b}. Note that considering other
numbers of components does not significantly improve the results.

\begin{figure}[h!]
\centering
\includegraphics[width=\figwidth]{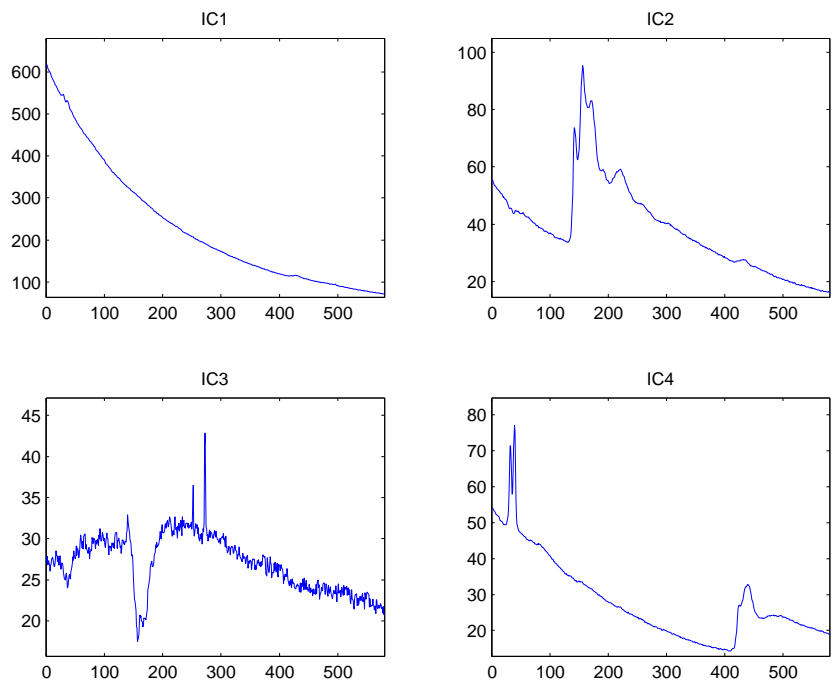}
\caption{Spectral components extracted by CubICA with a restricted
430-800 eV energy range. In this case the unmixing is quite
satisfactory but only 4 physical EELS spectra are identified. The
corresponding maps are in depicted Fig. \ref{fig:fig_9}.
\label{fig:fig_8}}
\end{figure}

\begin{figure}[h!]
\centering
\includegraphics[width=\figwidth]{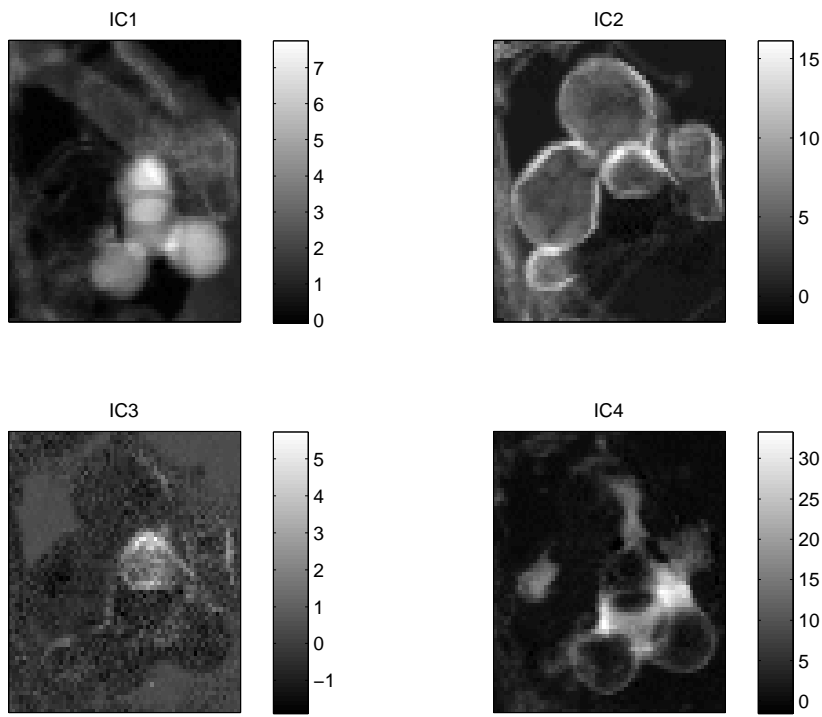}
\caption{Maps of the spectral components extracted by CubICA for a
limited energy range (the spectral signatures are depicted in Fig.
\ref{fig:fig_8}). \label{fig:fig_9}}
\end{figure}

We also tried to perform the analysis by restricting the energy
range to a window corresponding to B-K edge, i.e., 188-206 eV ,
following the strategy in \cite{Arenal2008}. However, once again,
ICA fails to unmix properly the components. By choosing the energy
range 330-610 eV, which only corresponds to Ca, N and O, satisfying
unmixing results can be obtained with $4$ components: background,
BN, B2O3 and pure boron (Fig. \ref{fig:fig_8} and \ref{fig:fig_9}).
In this case, the differences between the two orientations of h-BN
are too small to be detected on the N-K edge, providing only one
component for h-BN. Consequently, it seems that ICA performs better
with high energy ranges, as it was the case in \cite{delapena2011}
with a 430-800 eV energy window. According to \cite{delapena2011},
this could be explained by non-linear effects caused by multiple
scattering and by the variance of the C-K edge which is of the same
order of magnitude as the other signals. In the analyzed example,
since B-K is the edge of interest, the energy levels that contains
this non-linearity can not be removed from the analysis without
loosing crucial and discriminative information initially contained
in the data.

\subsection{Spectral mixture analysis with VCA and BLU}

SMA of the EELS spectrum-imaging data is conducted by using the BLU
algorithm presented in paragraph \ref{subsec:BLU}. We found that $R
= 8$ give the most satisfying results. The BLU algorithm has been
initialized with endmembers provided by the VCA algorithm introduced
in paragraph \ref{subsec:BLU}. Unmixing results provided by VCA are
also reported to be compared with endmembers identified by BLU. VCA
and BLU calculations were performed in the Matlab$^{\copyright}$
(Release 2010b) environment.

\begin{figure}[h!]
\centering
\includegraphics[width=\figwidth]{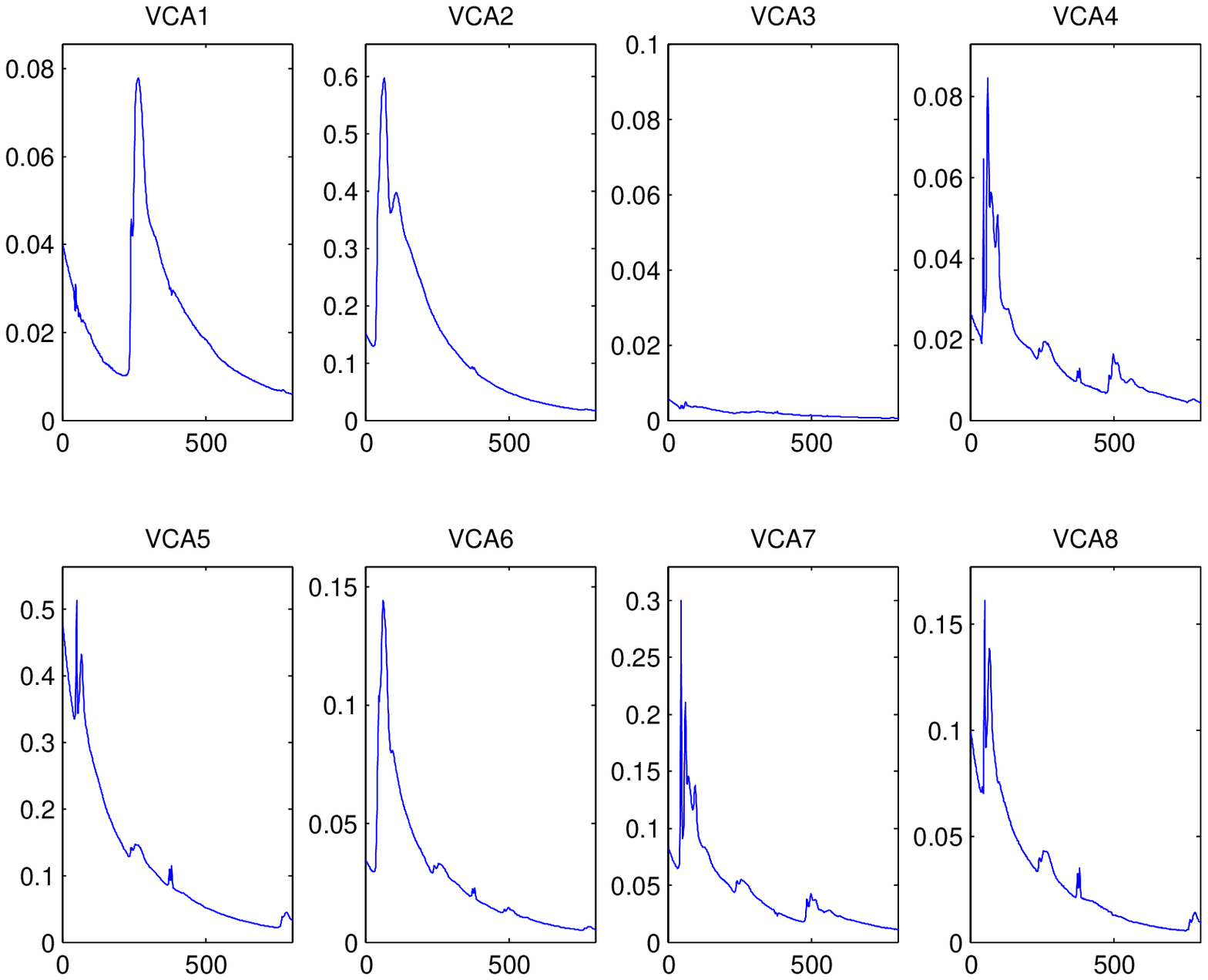}
\caption{Spectral components extracted by VCA. These endmembers can
be identified as real spectral, since they are chosen among the
observed pixels. The corresponding maps are depicted in Fig.
\ref{fig:fig_11}. \label{fig:fig_10}}
\end{figure}

\begin{figure}[h!]
\centering
\includegraphics[width=\figwidth]{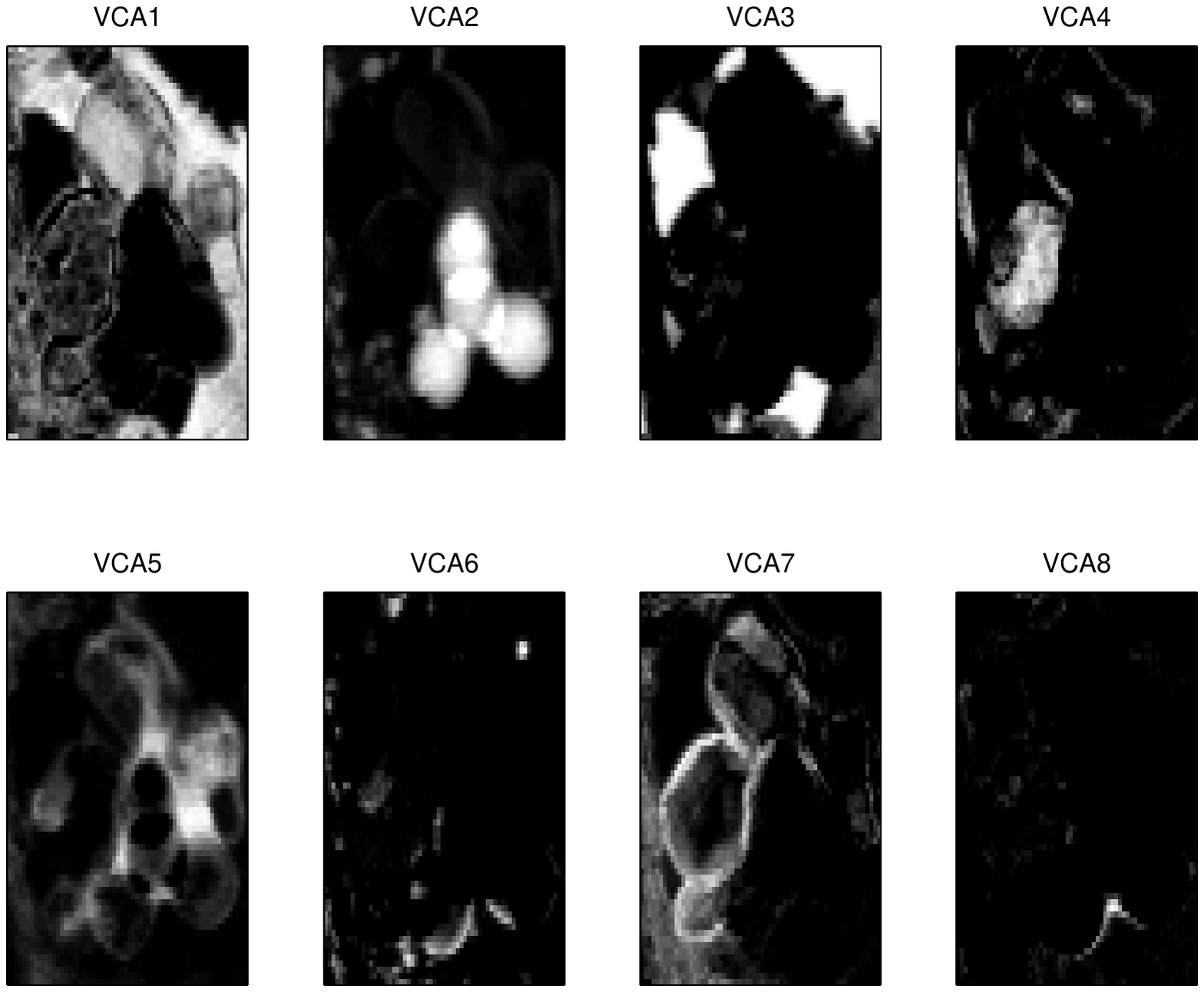}
\caption{Maps of the spectral components extracted by VCA (the
spectral signatures are depicted in Fig. \ref{fig:fig_10}).
\label{fig:fig_11}}
\end{figure}

Results obtained with VCA are presented in Fig. \ref{fig:fig_10} and
\ref{fig:fig_11}. It clearly appears that i) all spectra correspond
to realistic EELS spectra with characteristic edges on a decreasing
background, and ii) the related maps correctly separate different
areas on the sample, which was not the case for maps obtained with
ICA. The comparison of maps and endmembers with results obtained in
\cite{Arenal2008} allows some target signature to be easily
identified:
\begin{itemize}
  \item according to the C map of \cite{Arenal2008}, component VCA1 corresponds to the C supporting film;
  \item VCA2 and VCA6 both correspond to pure B in \cite{Arenal2008}. This is similar to
the case of AVIRIS hyperspectral data where the ``playa'' signature
is separated into two distinct regions \cite{Chang2004tgrs}. It is
likely that the splitting of the pure-B component does not
correspond to 2 physically distinct signals.
 \item VCA3 is related to holes in the
sample, thus there is no characteristic signal. This component is
nevertheless necessary to account for the absence of signal in these
pixels.
 \item VCA4 corresponds to B-N$^{\sigma*}$ but the map is slightly different
from the one obtained in \cite{Arenal2008}.
 \item VCA5 can be associated with B2O3
since fine structure in the corresponding abundance map the presence
of O are in good agreement with results obtained in
\cite{Arenal2008} for boron oxide.
 \item Component VCA7 correspond to B-N$^{\pi*}$, with a observable N-K
edge.
\end{itemize}

\begin{figure}[h!]
\centering
\includegraphics[width=\figwidth]{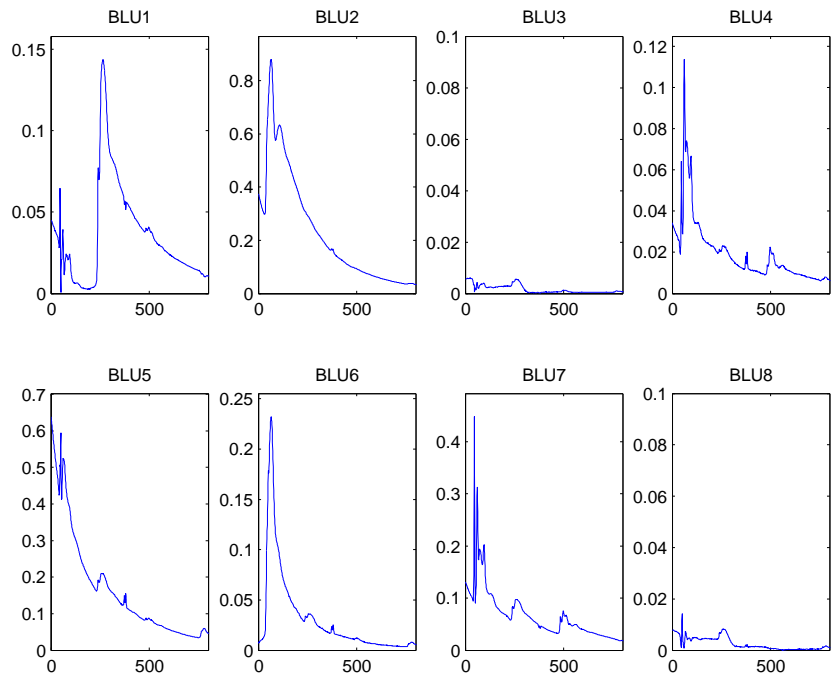}
\caption{Spectral components estimated by BLU. The recovered
endmembers properly correspond to EELS spectra. Contrary to VCA,
these signatures are not initially present in the measures EELS
spectrum-image. Indeed, BLU does not require the assumption of the
presence of pure pixels in the analyzed image. The corresponding
maps are depicted in Fig. \ref{fig:fig_13}. \label{fig:fig_12} }
\end{figure}

\begin{figure}[h!]
\centering
\includegraphics[width=\figwidth]{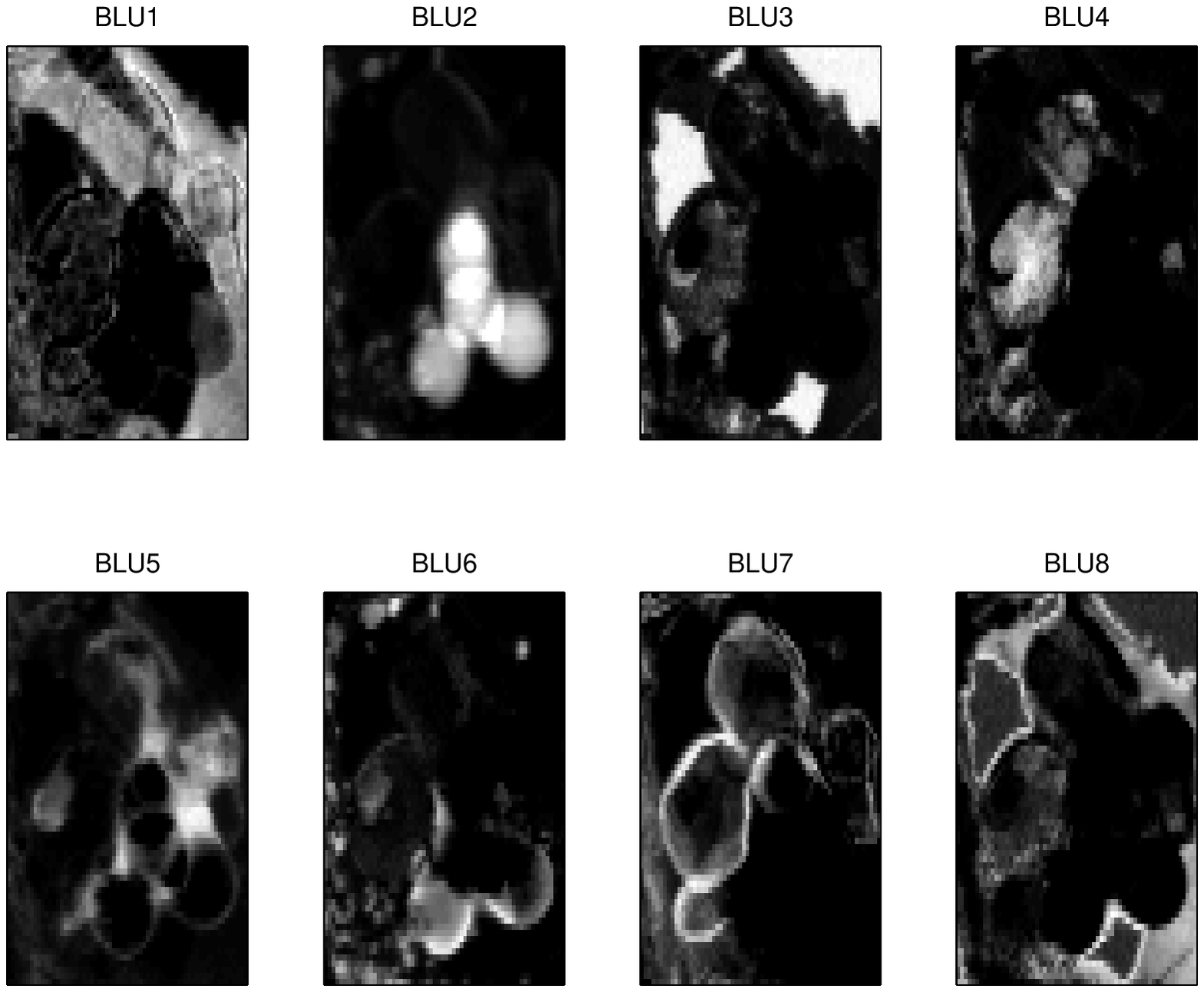}
\caption{Maps of the spectral components extracted by BLU (the
spectral signatures are depicted in Fig. \ref{fig:fig_12}).
\label{fig:fig_13} }
\end{figure}

The endmember spectra estimated by the proposed BLU algorithm are
depicted in Fig. \ref{fig:fig_12} and the abundance maps in Fig.
\ref{fig:fig_13}. For some components, results are quite similar to
those obtained with VCA:
\begin{itemize}
  \item VCA7 and BLU7 correspond to B-N$^{\pi*}$ with an identifiable N-K edge.
  \item VCA4 and BLU4 correspond to B-N$^{\sigma*}$
  \item  VCA1 and BLU1 correspond to the C supporting film.
  \item Pure B is separated into two components, BLU2 and
BLU6 (VCA2 and VCA6, respectively).
\end{itemize}

However some endmembers unmixed by BLU are significantly different:
\begin{itemize}
  \item Whereas B-O signature was divided into 2 distinct components with
VCA (VCA5 and VCA8), BLU is able to identify only one spectral
signature with a strong O signal (BLU5).
\item Vacuum signal is classified
into 2 components (BLU3 and BLU8).
\end{itemize}

This later feature is quite difficult to be interpreted. When
applied with only 7 components, the BLU algorithm does not separate
the components corresponding to B-N$^{\pi*}$ and B-N$^{\sigma*}$
although the vacuum signature is still decomposed into two distinct
signatures. Some authors report that some minor components can be
masked by the spectral variability of major components
\cite{Keshava2002,Chang2004tgrs,Chang2002tgrs}. It can be thus
necessary to consider a number of components greater that the number
of targets to be identified.

\begin{figure}[h!]
 \centering
\includegraphics[width=\figwidth]{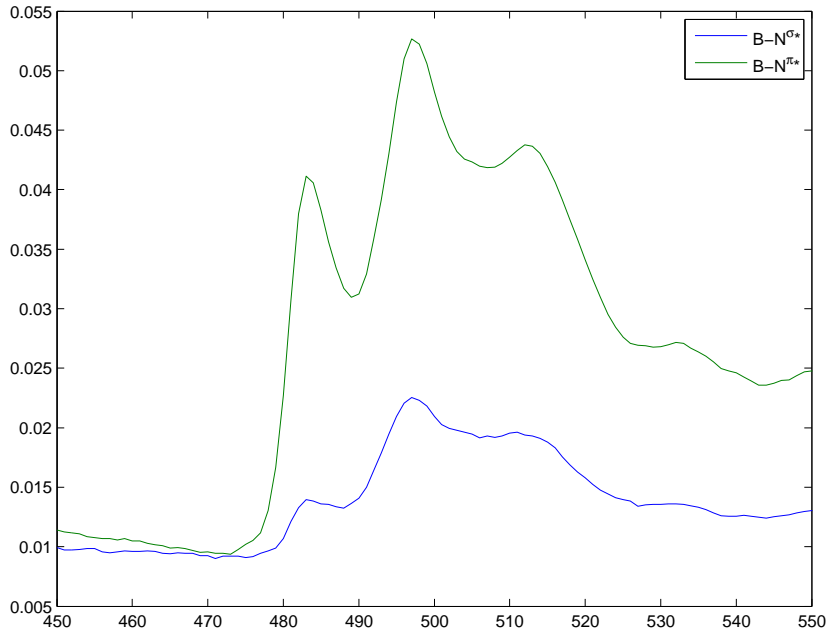}
\caption{Detail of the N K edge, with subtracted background and
scaled intensities, for endmembers BLU4 (B-N${\sigma*}$) and BLU7
(B-N$^{\pi*}$). \label{fig:fig_14}}
\end{figure}

Restricting the analysis to an energy window corresponding to B-K
edge does not improve significantly the results. Furthermore, when
considering only a restrictive part of the spectra, relevant
information composed of the different edges can be lost. For
instance, endmember BLU5 with a strong O-K edge is associated with a
B-K edge whose fine structure undoubtedly corresponds to B2O3.
Endmember BLU4 corresponding to a high ${\pi*} \slash {\sigma*}$
ratio for the B-K edge includes a N-K edge with the same feature
(Fig. \ref{fig:fig_14}).

The maps obtained with BLU seem to be in good agreement with those
presented in \cite{Arenal2008}, in particular with a higher
intensity of component BLU4 corresponding to particle 2 (particles
are located in Fig. \ref{fig:fig_2}). The small particle 3 is also
better defined with BLU7 than with VCA7. This better agreement of
the maps with the one found in \cite{Arenal2008} illustrate the
accuracy of the BLU method when conducting SMA.

\section{Discussion}

PCA is one of the most commonly used technique to identify
significant patterns from multivariate data. As the EELS signatures
to be recovered are not orthogonal, components recovered by PCA do
not have any physical meaning. As a consequence, it is quite
legitimate to conclude that PCA fails to perform interesting
spectral unmixing. Nevertheless, since the most relevant components
identified by PCA can be used to reconstruct the spectrum-image, PCA
can be advocated as a powerful filtering technique, e.g., to denoise
the measured signal. Traditional chemical mapping can then be
performed on the filtered spectrum-image with a significant increase
of the signal-to-noise ratio. However, to go further in the data
analysis, it is necessary to resort to more advanced analysis
methods.

In \cite{Arenal2008} bonding maps have been obtained by fitting
reference spectra manually extracted from regions of pure compounds.
Nevertheless, this supervised method requires a careful inspection
of both the elemental maps and the fine structure to correctly
select the reference spectra. Advantages of fully unsupervised
analysis such as SMA are to rely as little as possible on these
subjective choices operated by an expert. In addition, in certain
practical circumstances, these choices can be not straightforward.
For instance, the pure boron map of \cite{Arenal2008} actually
corresponds to 2 distinct components identified when conducting SMA
(BLU2 and BLU6). Consequently, in this typical case, it would be
difficult to decide which component should be chosen as a reference
for the least square fitting method employed in \cite{Arenal2008}.

In various application fields, ICA has been considered as an
efficient tool to extract sources from mixed signals. Plenty of
ICA-based methods have been proposed in the literature, and numerous
toolboxes are even freely available. These matrix factorization
techniques rely on the independence of the signatures to be
recovered. However, independence is rather a stringent condition in
the targeted application focused in this paper. Indeed, EELS
spectrum-images seldom fulfill this critical requirement.
Consequently, even if ICA has provided interesting results in some
specific cases \cite{delapena2011} \cite{Trasobares2011}, components
extracted by this methodology have been demonstrated to be
difficulty interpretable.

Contrary to PCA and ICA, SMA does not require any orthogonality or
independence assumptions on the components. Conversely, by
explicitly constraining the signatures to be non-negative and the
abundances to be related to proportions (i.e., with sum-to-one and
positivity constraints), SMA allows the interpretability of the
identified patterns to be guaranteed. The statistical BLU algorithm,
designed to perform SMA, was able to extract endmembers close to the
reference spectra manually extracted in \cite{Arenal2008}. Contrary
to VCA which is a geometrical unmixing method, BLU does not require
the presence of pure pixels in the analyzed spectrum-image, i.e.,
pixels composed of a unique endmember. Consequently, BLU has
demonstrated undeniable abilities to extract relevant components
from EELS spectrum image, and to provide an accurate mapping of
these components over the sample.

\section{Conclusions}

This work demonstrated the interest of using spectral unmixing,
initially devoted to remote sensing images, to perform fine
structure analysis of EELS spectrum-images. Several unmixing
methods, namely VCA and BLU, were presented as alternative analysis
methods to PCA, ICA or least square fitting. According to the
conducted study, VCA algorithm was noteworthy for its low
computational complexity and could be used on line for a first check
of the data during the STEM experiments . At a higher computational
price, BLU provided a finer and more relevant mapping of the
spectral components. In particular, obtained results were all the
more promising as the studied sample was rather complicated, with
the presence of vacuum, amorphous carbon support, contamination
unexpected elements as Ca.

Spectral mixture analysis, and more specifically the BLU algorithm,
represent a significant step in the evolution of the multivariate
analysis methods able to extract relevant information from EELS
data. More generally, SMA brings an efficient solution to the
crucial issue that consists of processing an increasing amount of
collected data -- in 1998 the data set consisted of only $64$
spectra \cite{Bonnet1999}, whereas spectrum images of $128\times128$
pixels are now frequently acquired. One of the main advantage of
spectral unmixing methodology is its ability of providing more
detailed and more interpretable information about the fine structure
of the edges. This work significantly widens the range of analysis
methodologies available for the EELS community.

\section*{Acknowledgments}

The authors would like to thank O. Stephan and R. Arenal for
providing the EELS data.

\bibliographystyle{elsarticle-num}
\bibliography{Y:/all_ref}

\begin{thebibliography}{10}
\expandafter\ifx\csname url\endcsname\relax
  \def\url#1{\texttt{#1}}\fi
\expandafter\ifx\csname urlprefix\endcsname\relax\def\urlprefix{URL }\fi
\expandafter\ifx\csname href\endcsname\relax
  \def\href#1#2{#2} \def\path#1{#1}\fi

\bibitem{Colliex1994}
C.~Colliex, M.~Tencé, E.~Lefèvre, C.~Mory, H.~Gu, D.~Bouchet, C.~Jeanguillaume,
  Electron energy loss spectrometry mapping, Microchimica Acta 114 (1994)
  71--87.

\bibitem{Pennycook2011}
T.~J. Pennycook, M.~P. Oxley, J.~Garcia-Barriocanal, F.~Y. Bruno, C.~Leon,
  J.~Santamaria, S.~T. Pantelides, M.~Varela, S.~J. Pennycook, Seeing oxygen
  disorder in {YSZ/SrTiO3} colossal ionic conductor heterostructures using
  {EELS}, The European Physical Journal - Applied Physics 54~(03).

\bibitem{Bonnet1999}
N.~Bonnet, N.~Brun, C.~Colliex, Extracting information from sequences of
  spatially resolved {EELS} spectra using multivariate statistical analysis,
  Ultramicroscopy 77~(3--4) (1999) 97--112.

\bibitem{HREM}
\href{http://www.hremresearch.com/}{The {HREM Research Inc.} website}.
\newline\urlprefix\url{http://www.hremresearch.com/}

\bibitem{Bosman2006}
M.~Bosman, M.~Watanabe, D.~Alexander, V.~Keast, Mapping chemical and bonding
  information using multivariate analysis of electron energy-loss spectrum
  images, Ultramicroscopy 106~(11--12) (2006) 1024--1032.

\bibitem{Bonnet2005}
N.~Bonnet, D.~Nuzillard, Independent component analysis: A new possibility for
  analysing series of electron energy loss spectra, Ultramicroscopy 102~(4)
  (2005) 327--337.

\bibitem{delapena2011}
F.~de~la Pe\~na, M.-H. Berger, J.-F. Hochepied, F.~Dynys, O.~Stephan, M.~Walls,
  Mapping titanium and tin oxide phases using {EELS}: An application of
  independent component analysis, Ultramicroscopy 111~(2) (2011) 169--176.

\bibitem{Guiton2011}
B.~S. Guiton, V.~Iberi, S.~Li, D.~N. Leonard, C.~M. Parish, P.~G. Kotula,
  M.~Varela, G.~C. Schatz, S.~J. Pennycook, J.~P. Camden, Correlated optical
  measurements and plasmon mapping of silver nanorods, Nano Letters 11~(8)
  (2011) 3482--3488.

\bibitem{Keenan2009}
M.~R. Keenan, Exploiting spatial-domain simplicity in spectral image analysis,
  Surface and Interface Analysis 41~(2) (2009) 79--87.

\bibitem{Keshava2002}
N.~Keshava, J.~F. Mustard, Spectral unmixing, IEEE Signal Processing Magazine
  (2002) 44--57.

\bibitem{Bioucas2012jstars}
J.~M. Bioucas-Dias, A.~Plaza, N.~Dobigeon, M.~Parente, Q.~Du, P.~Gader,
  J.~Chanussot, Hyperspectral unmixing overview: Geometrical, statistical, and
  sparse regression-based approaches, IEEE J. Sel. Topics Appl. Earth
  Observations Remote Sens. 5~(2) (2012) 354--379.

\bibitem{Dobigeon2009}
N.~Dobigeon, S.~Moussaoui, M.~Coulon, J.-Y. Tourneret, A.~O. Hero,
  \href{{http://dobigeon.perso.enseeiht.fr/app\_hyper\_SMA.html}}{Joint
  {B}ayesian endmember extraction and linear unmixing for hyperspectral
  imagery}, IEEE Trans. Signal Processing 57~(11) (2009) 4355--4368.
\newline\urlprefix\url{{http://dobigeon.perso.enseeiht.fr/app\_hyper\_SMA.html%
}}

\bibitem{Nascimento2005}
J.~M. Nascimento, J.~M. {Bioucas-Dias}, Vertex component analysis: a fast
  algorithm to unmix hyperspectral data, IEEE Trans. Geosci. and Remote Sensing
  43~(4) (2005) 898--910.

\bibitem{Arenal2008}
R.~Arenal, F.~de~la Pe\~na, O.~St\'ephan, M.~Walls, M.~Tenc\'e, A.~Loiseau,
  C.~Colliex, Extending the analysis of {EELS} spectrum-imaging data, from
  elemental to bond mapping in complex nanostructures, Ultramicroscopy 109~(1)
  (2008) 32--38.

\bibitem{Chang2004tgrs}
C.-I. Chang, Q.~Du, Estimation of number of spectrally distinct signal sources
  in hyperspectral imagery, IEEE Trans. Geosci. and Remote Sensing 42~(3)
  (2004) 608--619.

\bibitem{Nascimento2005b}
J.~M.~P. Nascimento, J.~M. {Bioucas-Dias}, Does independent component analysis
  play a role in unmixing hyperspectral data?, IEEE Trans. Geosci. and Remote
  Sensing 43~(1) (2005) 175--187.

\bibitem{Chang2002tgrs}
C.-I. Chang, S.-S. Chiang, J.~Smith, I.~Ginsberg, Linear spectral random
  mixture analysis for hyperspectral imagery, IEEE Trans. Geosci. and Remote
  Sensing 40~(2) (2002) 375--392.

\bibitem{Trasobares2011}
S.~Trasobares, M.~L\'opez-Haro, M.~Kociak, K.~March, F.~de~La~Pe\~na, J.~A.
  Perez-Omil, J.~J. Calvino, N.~R. Lugg, A.~J. D'Alfonso, L.~J. Allen,
  C.~Colliex, Chemical imaging at atomic resolution as a technique to refine
  the local structure of nanocrystals, Angewandte Chemie International Edition
  50~(4) (2011) 868--872.

\end{thebibliography}

\end{document}